\documentclass[aps,prc,twocolumn,showpacs,nofootinbib]{revtex4-1}
\usepackage[utf8]{inputenc}
\usepackage{graphicx}   
\usepackage{latexsym}   
\usepackage{enumerate}
\usepackage{rotating,booktabs,multirow}
\usepackage{amsmath}
\usepackage{amsfonts}
\usepackage{amssymb}
\usepackage{multirow}
\usepackage{bm}
\usepackage{xcolor}

\begin{document}
\title{Effects of a phase transition on two-pion interferometry in heavy-ion collisions at $\sqrt{s_\mathrm{NN}}=2.4 - 7.7$ GeV}
\author{Pengcheng~Li$^{1,2,3}$, Jan Steinheimer$^{4}$, Tom~Reichert$^{1,6}$, Apiwit~Kittiratpattana$^{1,7}$, Marcus~Bleicher$^{1,5,6}$,Qingfeng Li$^{2,3}$\footnote{liqf@zjhu.edu.cn}}

\affiliation{$^1$ Institut f\"ur Theoretische Physik, Goethe Universit\"at Frankfurt, Max-von-Laue-Strasse 1, D-60438 Frankfurt am Main, Germany}
\affiliation{$^2$ School of Science, Huzhou University, Huzhou 313000, China}
\affiliation{$^3$ School of Nuclear Science and Technology, Lanzhou University, Lanzhou 730000, China}
\affiliation{$^4$ Frankfurt Institute for Advanced Studies, Ruth-Moufang-Str. 1, D-60438 Frankfurt am Main, Germany}
\affiliation{$^5$ GSI Helmholtzzentrum f\"ur Schwerionenforschung GmbH, Planckstr. 1, 64291 Darmstadt, Germany}
\affiliation{$^6$ Helmholtz Research Academy Hesse for FAIR (HFHF), GSI Helmholtz Center for Heavy Ion Physics, Campus Frankfurt, Max-von-Laue-Str. 12, 60438 Frankfurt, Germany}
\affiliation{$^7$ Suranaree University of Technology, University Avenue 111, Nakhon Ratchasima 30000, Thailand}

\begin{abstract}
Hanbury-Brown-Twiss (HBT) correlations for charged pions in central Au+Au collisions at $\sqrt{s_\mathrm{NN}}=2.4 - 7.7$ GeV (corresponding to beam kinetic energies in the fixed target frame from $E_{\rm{lab}}=1.23$ to 30 GeV/nucleon) are calculated using the UrQMD model with different equations of state.
The effects of a phase transition at high baryon densities is clearly observed in the HBT parameters that are explored.
It is found that the available data on the HBT radii, $R_{O}/R_{S}$ and $R^{2}_{O}-R^{2}_{S}$, in the investigated energy region favors a relatively stiff equation of state at low beam energies which then turns into a soft equation of state at high collision energies consistent with astrophysical constraints on the high density equation of state of QCD.
The specific effects of two different phase transition scenarios on the $R_{O}/R_{S}$ and $R^{2}_{O}-R^{2}_{S}$ are investigated.
It is found that a phase transition with a significant softening of the equation of state below 4 times nuclear saturation density can be excluded using HBT data.
Our results highlight that the pion's $R_{O}/R_{S}$ and $R^{2}_{O}-R^{2}_{S}$ are sensitive to the stiffness of the equation of state, and can be used to constrain and understand the QCD equation of state in the high baryon density region.
\end{abstract}

\maketitle

\section{Introduction}\label{section1}
The exploration of the properties of hot and dense nuclear matter is among the major goals of today's largest accelerator facilitates.
Theoretically, such matter is described by the theory of strong interaction, called Quantum-Chromo-Dynamics (QCD). To obtain ab-initio results of QCD, one is unfortunately restricted to lattice QCD calculations for static systems at high temperatures and small baryo-chemical potentials (the main reason for this restriction is the so-called sign-problem \cite{Pandav:2022xxx}).
The current state-of-the-art lattice QCD calculations predict a crossover transition between the hadronic phase and quark-gluon plasma (QGP) phase at vanishing baryon density at a temperature of $T\simeq$ 150 MeV \cite{Aoki:2006we,Bazavov:2011nk}.
Many model calculations predict that with the increase of net-baryon density, the phase transition becomes first order and ends at a critical endpoint (CEP) at finite temperature and density \cite{Braun-Munzinger:2015hba,Bzdak:2019pkr,Gunkel:2021oya}.
Relativistic heavy-ion collisions (HICs) at terrestrial laboratories allow to investigate the properties of strongly interacting matter in a controlled environment.
By changing the mass or centrality of the impinging nuclei and the collision energy, one can vary the initially created densities and temperatures, which lead after an approximately isentropic expansions to different freeze-out conditions of baryon chemical potential ($\mu_{B}$) and temperature $T$ \cite{Bzdak:2019pkr}.

In order to investigate signatures of a deconfined QGP, and search for evidence of a possible first-order phase transition and the location of its CEP, on the experimental side, several experimental programs at  GSI, BNL, and CERN have been successfully run.
To obtain further data future facilities like FAIR, NICA and HIAF are proposed and currently build.
Over the past decades, in the first phase of the Beam Energy Scan program at RHIC (BES-\uppercase\expandafter{\romannumeral1}), the Au+Au collisions data at $\sqrt{s_{\rm{NN}}}=7.7$ to 200 GeV were collected and analysed.
According to the results from BES-\uppercase\expandafter{\romannumeral1},
the region of interest can be narrowed to collision energies below $\sqrt{s_{\rm{NN}}}=20~\text{GeV}$ \cite{starnote0598}.
Since the lowest beam energy which is accessible at RHIC in collider mode is $\sqrt{s_{\rm{NN}}}=7.7~\text{GeV}$, a fixed-target (FXT) program has been developed to allow the STAR experiment to access energies from $\sqrt{s_{\rm{NN}}}=3.0$ to 7.7 $\text{GeV}$ \cite{STAR:2002eio}.

In addition to this large body of (upcoming) experimental data, there have been substantial developments on the theoretical and modelling side.
As an important tool to extract information on the nuclear equation of state (EoS) and the properties of hadrons from low to relativistic-energy HICs, transport theories have been used for many years.
To establish a theoretical systematic error and disentangle the causes that lead to different predictions, various comparisons of different transport models have been performed over the years \cite{Bratkovskaya:2004kv,Bratkovskaya:2013vx,Reichert:2021ljd,TMEP:2022xjg,Bleicher:2022kcu}.
Also, many hydrodynamic approaches and hybrid models, which incorporate different EoS, have been widely used to understand the properties of dense strongly interacting matter at ultra-relativistic beam energies \cite{Bass:1999tu,Dumitru:1999sf,Huovinen:2006jp,Steinheimer:2007iy,Petersen:2008dd,Steinheimer:2011ea,Shen:2020mgh}.
Various observables have been suggested to explore the locations of first order quark-hadron phase transition boundary and CEP, such as high-order cumulants \cite{STAR:2020tga,STAR:2021fge}, intermittency analysis \cite{NA49:2012ebu}, the yield ratio of light nuclei \cite{Sun:2018jhg}, Hanbury Brown and Twiss (HBT) interferometry \cite{Lacey:2014wqa} (see \cite{Bzdak:2019pkr,Bluhm:2020mpc} for an overview and references therein).

In this work, we will mainly focus on the pion intensity interferometry (HBT interferometry) \cite{HanburyBrown:1956bqd,HanburyBrown:1954amm}, which can be used to reveal the space-time substructure and momentum correlations of the freeze-out configuration in HICs.
For a detailed descriptions of the history and development of HBT interferometry, the reader is referred to Refs. \cite{HanburyBrown:1956bqd,HanburyBrown:1954amm,gglp,Pratt:1984su,Zajc:1984vb,Pratt:1986cc,Lisa:2005dd,E895:2000bxp,Retiere:2003kf,STAR:2014shf}.
It was argued that the HBT radii (source radii) parameters are sensitive to a first-order phase transition and may reveal the CEP in the QCD phase diagram \cite{Pratt:1986cc,Bertsch:1988db,Rischke:1996em,Li:2008qm}.
It was predicted that a nonmonotonic behaviour (maximum) in the excitation functions for the emission source radii ratio and difference obtained from two pion interferometry measurements in Au+Au collisions would serve as signal for a phase transition.

Such a behaviour is osberved by the STAR experiment \cite{Lacey:2014wqa,STAR:2014shf}, however, at a very high beam energy of $\sqrt{s_{\rm NN}}\approx 20~\text{GeV}$. Thus, the investigations about the effects of EoS on the HBT interferometry within different models are mostly restricted to high energies \cite{Li:2008qm,Zhang:2017axr,Batyuk:2017smw,Alqahtani:2020daq}.
Since other observables like e.g. fluctuations did not show the behaviour expected from a phase transition at such high beam energies, the interpretation is still in question.

To approach this challenge it is important to develop models which are able to not only predict single observables but offer the ability to predict a wide range of observables in a consistent way that allows direct comparison with experiments.
In a previous work \cite{OmanaKuttan:2022the} it was shown how any density dependent equation of state can be implemented in the microscopic transport model UrQMD.
This now offers the opportunity to implement different phase transition scenarios in a consistent way and study a variety of possible observables. Ultimately this will allow us to make consistent statements on the existence of a phase transition and its possible location.
Effects of a phase transition on hadronic flow observables have already been shown to be significant in this implementation of the UrQMD model \cite{Steinheimer:2022gqb}.
We also expect that the density dependent potentials, will make a sizable contribution to the proton and net-charge number fluctuations \cite{Steinheimer:2018rnd,Ye:2020lrc} during the collision.

Thus, it is interesting and necessary to explore the influence of the EoS on the pion interferometry in HICs at several GeV beam energies, within the same framework, to make predictions on concerted signals for the phase transition from different measurements.

This paper is organized as follows: in Sec. \ref{sec:2}, the UrQMD model and the methods are briefly described.
In Sec. \ref{sec:3} the three-dimensional pion HBT radius results are shown.
Finally, the conclusions are presented in Sec. \ref{sec:4}.

\section{Model and method}\label{sec:2}
\subsection{UrQMD model and EoS}
In the present study, we use the current version of the UrQMD transport model (UrQMD 3.5) \cite{Bass:1998ca,Bleicher:1999xi,Bleicher:2022kcu} to investigate the pion intensity interferometry in heavy-ion collisions.
The UrQMD model can be applied in different modes. At high energies, the cascade mode in which the hadrons interact through binary scattering according to a geometrical interpretation of elastic and inelastic cross sections is most often used.
At lower energies, it is also necessary to incorporate the nuclear interactions for a complete modeling of the transport dynamics (calculation with nuclear potential).
In the mode when nuclear potential interactions are taken into account, each hadron is represented by Gaussian wave packets with a certain width, and after the initialization of projectile and target nuclei, the position and momentum of the {\it i}-th hadron is propagated according to Hamilton's equation of motion, which read as: $\dot{\textbf{r}}_{i}=\frac{\partial  \langle H  \rangle}{\partial\textbf{p}_{i}},~ \dot{\textbf{p}}_{i}=-\frac{\partial  \langle H \rangle}{\partial \textbf{r}_{i}}.$
Here, {\it $\langle H \rangle$} is the total Hamiltonian function of the system, it consists of the kinetic energies $\sum_i T_i$ and the effective interaction potential energies $\sum_i V_i$ of all baryons $i$ in the system.
In the default version of the UrQMD model, the potential energies include the two-body and three-body Skyrme-, Yukawa-, Coulomb-, and Pauli-terms \cite{Li:2010ie,Hillmann:2018nmd,Hillmann:2019wlt,Li:2005gfa} \footnote{The Pauli-term is usually turned off and Yukawa-term is negligible for the beam energies investigated here.} .
The Skyrme potential is computed from the single particle energy as ${U(\rho_b)=\frac{\partial (\rho_b  \cdot V(\rho_b))}{\partial \rho_b}}$ and can be expressed as $U=\alpha\left(\frac{\rho_{b}}{\rho_{0}}\right)+\beta\left(\frac{\rho_{b}}{\rho_{0}}\right)^{\gamma}$, where $\rho_{b}$ is the baryonic interaction density in units of $\rho_{0}=0.16$ fm$^{-3}$, the ground state baryon density.
By changing the parameters $\alpha$, $\beta$, and $\gamma$, one can change the stiffness  of EoS (usually termed a hard or soft EoS for a large or small value of the incompressibility $K_0$).
In the following, calculations within the cascade mode, representing a hadron resonance gas EoS without potentials \cite{Petersen:2008dd,Steinheimer:2009nn,Motornenko:2019arp} and a hard Skyrme EoS ($K_0=380$ MeV) in the molecular dynamics mode will serve as benchmark simulations \cite{Hillmann:2019wlt}.

In addition, several novel equations of state are implemented through effective density dependent potentials to gain insights into the properties of strong-interacting matter. Here, the EoS is based on a realistic chiral mean field (CMF) model with different phase transition scenarios, adopted to explore the sensitivity of the pion interferometry to the EoS.
This CMF EoS was first incorporated in the UrQMD model in Ref. \cite{OmanaKuttan:2022the}, and it was achieved by devising a method by which the mean field potential energy $V$ that enters the equations of motion can be calculated from the energy per baryon of the CMF model \cite{Motornenko:2019arp} and which is described in detail in Ref. \cite{OmanaKuttan:2022the}.
The CMF model incorporates the main concepts of QCD phenomenology: chiral interactions in the baryon octet, the full PDG hadron list, excluded volume repulsive interactions among all hadrons, baryon parity doubling, and quarks coupled to an effective Polyakov loop potential.
The CMF model  describes many aspects of QCD phenomenology, and has been widely employed as EoS in the hydrodynamic simulations of both heavy ion collisions and binary neutron star mergers. The detailed description of the CMF model and applications to the exploration of heavy ion collisions can be found in Refs. \cite{Steinheimer:2010ib,Steinheimer:2011ea,Mukherjee:2016nhb,Motornenko:2019arp,Motornenko:2020yme,OmanaKuttan:2022the,Jakobus:2022ucs,Most:2022wgo,Seck:2020qbx}.
In further studies it was shown how one can extend this formalism to include a phase transition at high density \cite{Steinheimer:2022gqb}.
This phase transition is characterized by an unstable region, i.e. a range in density at which the isothermal speed of sound becomes imaginary.
In addition to the phase transition scenario (PT2) which was already introduced in Ref. \cite{Steinheimer:2022gqb} we now also include a new phase transition scenario PT3 and which we will discuss below.

\begin{figure}[t]
\begin{centering}
\includegraphics[width=0.43\textwidth]{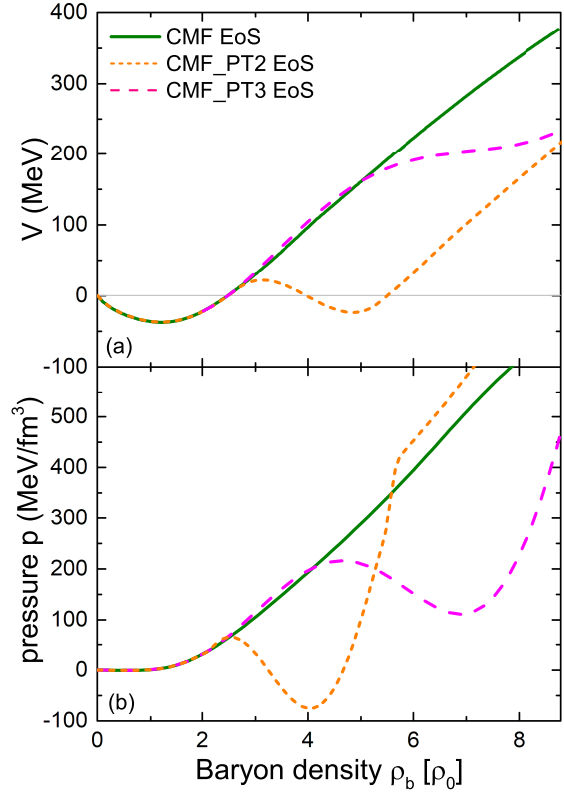}
\caption {\label{fig1}(Color online) The density dependent potential field energy $V$ [panel (a)] and corresponding pressure $p$ [panel (b)] shown for the different scenarios used in the CMF EoS. CMF$\_$PT2 and CMF$\_$PT3 both include a phase transition and unstable region at different densities, indicated by the negative slope of the pressure with respect to the density. The default CMF EoS corresponds to a smooth crossover transition.}
\end{centering}
\end{figure}

Fig. \ref{fig1} (a) shows the effective field energy per baryon calculated from different scenarios of the CMF EoS.
A direct comparison to the hard Skyrme potential EoS can be found in Ref. \cite{OmanaKuttan:2022the}. The standard CMF EoS shows a similar behavior in the mean field potential $V$ as that of the hard Skyrme EoS above saturation up to about 4 times saturation density ($\rho_{0}$), and then softens at even higher densities due to a crossover to the high density limit of a free gas of three quark flavors.
All three kinds of CMF EoS show the same behaviour in $V$ and $p$ up to about 3$\rho_{0}$.

As it is more instructive to discuss the properties of the EoS in terms of the pressure, panel (b) of Fig.\ref{fig1} therefore shows the pressure of the effective EoS as function of the baryon density. The pressure is calculated by
\begin{equation}
P(\rho_b,T) = P^{\rm id}(\rho_b,T) + \int_0^{\rho_b} \rho' \frac{\partial U(\rho')}{\partial \rho'} d\rho',
\end{equation}
where $P^{\rm id}(\rho_b,T)$ is the pressure of an ideal fermi-gas of hadrons and $U(\rho_b)$ is the single particle energy per baryon.

As the density increases, the CMF$\_$PT2 EoS becomes mechanically unstable at 2.5 times saturation density and reaches the minimum of the pressure at about 4$\rho_{0}$.
The CMF$\_$PT3 EoS behaves similarly to the standard CMF EoS until about 5$\rho_{0}$, above which it then becomes also mechanically unstable due to the phase transition.
This means while in the PT2 case the transition can be reached already at a rather low density (low collision energies), the onset of the unstable phase in PT3 is at a density which might be difficult to reach for most collisions systems.

\subsection{Pion HBT analysis}
To explore the effects of the various EoS with and without the different phase transitions, we perform UrQMD calculations to obtain the pions' freeze-out phase space coordinates.
Freeze-out in UrQMD is defined as the space-time point of last interaction (either a collision or a decay).
The freeze-out space-time coordinates and 4-momenta serve as  input for  the ``correlation after-burner" (CRAB v3.0$\beta$) \cite{CRAB} program, provided by S. Pratt. CRAB constructs the HBT correlation function defined as:
\begin{equation}\label{cpq}
C(\textbf{k},\textbf{q})=1+\frac{\int d^4x_1 d^4x_2 S(x_1,\textbf{p}_1) S(x_2,\textbf{p}_2) |\phi(\textbf{q}, \textbf{r})|^2} {{\int d^4x_1 S(x_1,\textbf{p}_1)}{\int d^4x_2 S(x_2,\textbf{p}_2)}}.
\end{equation}
Here, $\textbf{q}=\textbf{p}_1-\textbf{p}_2$ and $\textbf{k}=(\textbf{p}_{1}+\textbf{p}_{2})/2$ are the relative momentum and the average momentum of the two particles.
$S(x,\textbf{p})$ represents the probability for emitting a particle with momentum $\textbf{p}$ from the space-time point $x=(\textbf{r}, t)$.
$\phi(\textbf{q}, \textbf{r})$ is the relative two-particle wave function with $\textbf{r}$ being their relative position.

The correlation function is then fitted assuming a three-dimensional Gaussian form in the longitudinally comoving system, which is expressed as
\begin{eqnarray}\label{f3d}
C(q_L,q_O,q_S)&&=N[(1-\lambda)+\lambda K_{C}(q_{inv},R_{inv})(1+\nonumber \\
&&\text{exp}(-R_L^2q_L^2-R_O^2q_O^2-R_S^2q_S^2-2R_{OL}^2q_Oq_L))],
\end{eqnarray}
where $N$ is the overall normalization factor, and $\lambda$ is the incoherence factor and lies between 0 (complete coherence) and 1 (complete incoherence) for bosons in realistic HICs \cite{Wong:2007hx}.
$K_{C}$ is the Coulomb correction factor depending on $q_{inv}$ and $R_{inv}$ \cite{Sinyukov:1998fc,HADES:2018gop,HADES:2019lek}, and  $q_{inv}=\frac{1}{2}\sqrt{({\bm p}_1 - {\bm p}_2)^2-(E_1-E_2)^2}$ is the invariant momentum.
The resulting HBT radii are $R_L$, $R_O$, and $R_S$ corresponding to the longitudinal (the beam direction), outward (the direction of the transverse component of the pair-momentum $\textbf{k}_{T}=(\textbf{p}_{1T}+\textbf{p}_{2T})/2$, and sideward directions (the direction is defined to be perpendicular to the other two directions), $R_{OL}$ is the cross-term, and $q_i$ is the pair relative momentum in the $i$ direction, such as, $q_{L}$ represents the pair relative momentum in the longitudinal direction.

\section{Results and discussions}\label{sec:3}
\subsection{Effects of the EoS without phase transitions}

\begin{figure}[ht]
\begin{centering}
\includegraphics[width=0.5\textwidth]{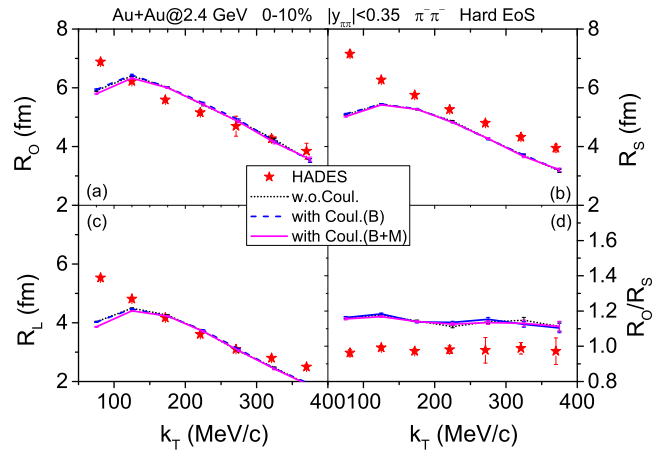}
\caption {\label{fig2}(Color online) $k_T$ dependence of pion HBT radii $R_O$ [panel (a)], $R_S$ [(b)], and $R_L$ [(c)], as well as the ratio $R_O/R_S$ [panel (d)] of $\pi^{-}$ source from central ($0-10\%$) Au+Au collisions at $\sqrt{s_{\rm{NN}}}=2.4$ GeV.
The results without Coulomb potential, with Coulomb potential for baryons only, and with the full Coulomb potential for all hadron are shown by dotted black, dashed blue and solid pink lines, separately. The solid stars represent the experimental data taken from Ref. \cite{HADES:2018gop}.}
\end{centering}
\end{figure}

To set the stage for the investigation of the influence of the different EoS, we will start with a discussion of the contribution of the Coulomb interaction on HBT observables using the standard hard Skyrme EoS, which also allows to describe flow data in this energy regime.
The contribution of the Coulomb potential on the HBT radii of negative pion pairs in Au+Au collisions at $\sqrt{s_{\rm{NN}}}=2.4$ GeV ($E_{\rm{lab}}=1.23$ GeV/nucleon) is shown in Fig. \ref{fig2}.
The calculations without Coulomb potential and with only the baryonic contribution to the Coulomb potential are shown by the dotted black and dashed blue lines, respectively. The calculations with the full Coulomb potential including also the meson are shown by the solid pink lines. The solid stars represent the experimental data taken from Ref. \cite{HADES:2018gop}.
The pair-rapidity is defined as $y_{\pi\pi}=\frac{1}{2}{\rm log}\left(\frac{E_1+E_2+p_{l 1}+p_{l 2}}{E_1+E_2-p_{l 1}-p_{l 2}}\right)$ is the two-pion rapidity with energies $E_1$ and $E_2$ and longitudinal momenta $p_{l 1}$ and $p_{l 2}$ in the center of mass system is chosen.
Here we employ a cut of $|y_{\pi\pi}|<$ 0.35 in line with the data.

\begin{figure*}[ht]
\centering
\includegraphics[width=0.82\textwidth]{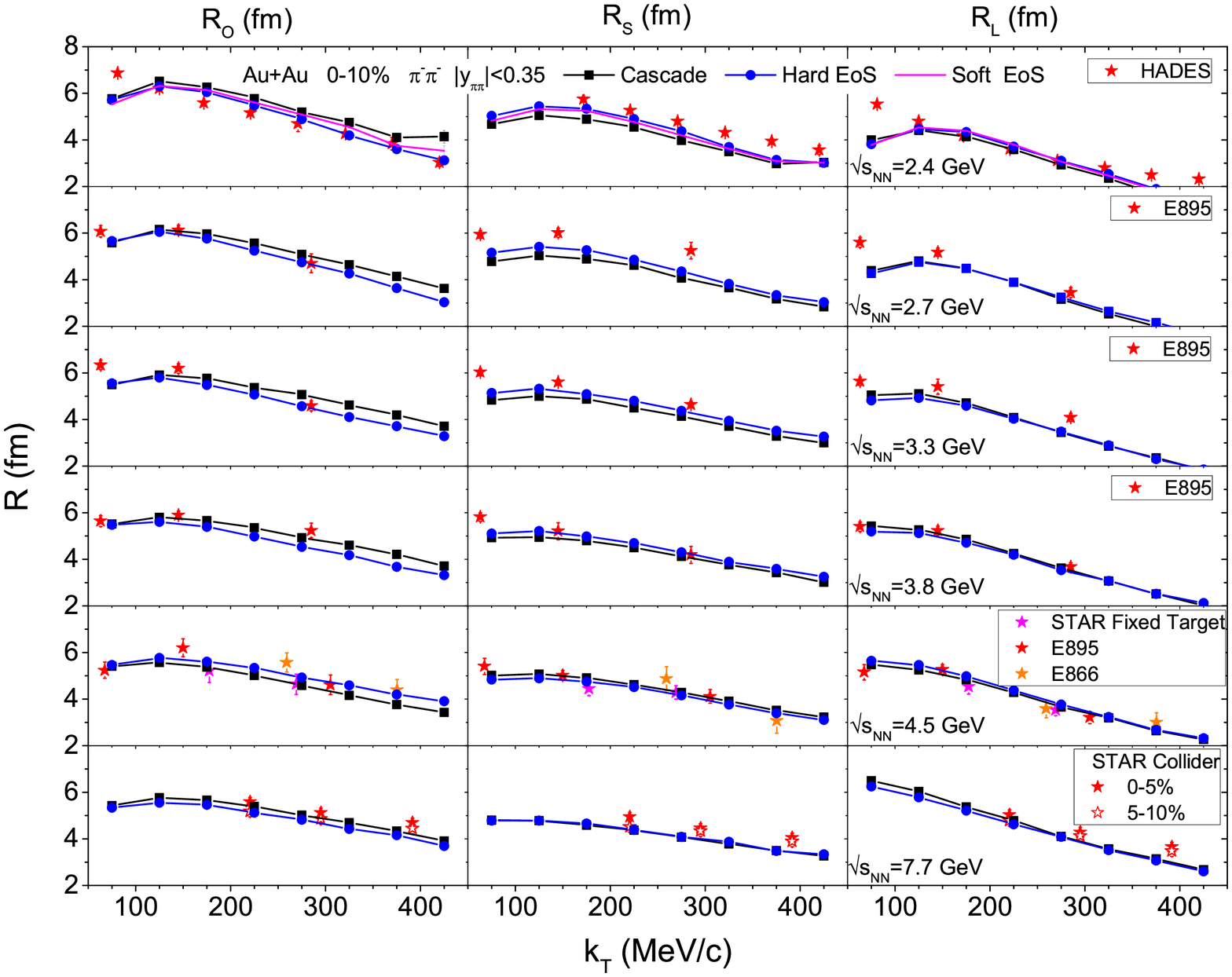}
\caption {\label{fig3}(Color online) Transverse momentum, $k_T$, dependence of the HBT radii $R_O$ (left panels), $R_S$ (middle panels) and $R_L$ (right panels) for 0-10\% central Au+Au collisions at $\sqrt{s_\mathrm{NN}}=2.4 - 7.7$ GeV (from top to bottom).
The data are indicated by stars, which are obtained by the HADES, E895, E866, and STAR collaborations \cite{Lisa:2005dd,HADES:2018gop,HADES:2019lek,E895:2000bxp,Okorokov:2013fya,STAR:2014shf,STAR:2020dav}.
The calculations with cascade mode are shown by lines with black squares, and the calculations with potentials are shown by the blue lines with circles (hard EoS) and the pink lines (soft EoS).}
\end{figure*}
Firstly, it can be seen that all three calculations can reproduce the transverse momentum, $k_T$ ($\textbf{k}_T=(\textbf{p}_{1T}+\textbf{p}_{2T})/2$), dependence of the HBT radii $R_L$ and $R_O$, except for very small $k_T$ values.
The $R_S$ values in the calculations show slightly values smaller than the experimental data, which in turn makes the ratio $R_O/R_S$ (bottom right) larger than the values obtained from the experimental data.
By comparing the results calculated with and without Coulomb potential, one observes that the effects of the two-body mesonic Coulomb potential during the evolution on the HBT radii and the ratio $R_O/R_S$ are very weak.
This is different from flow observables, as was shown in a previous work which indicated that the yield and the collective flows of charged pions are indeed influenced by the mesonic Coulomb potential \cite{Li:2005gfa}.
Such a behaviour is to be expected, because the flow reflects the integrated collective motion of single particles, and therefore the Coulomb potential during the systems evolution does affect the (azimuthal) distribution of the pion emission more strongly.
Nevertheless, the size of the pion freeze-out source, as seen through the two-pion correlation function in relative momentum, is only very weakly affected.
Therefore, for the following discussion, we will omit the Coulomb effect on the HBT radii and focus only on the EoS dependence.

Fig. \ref{fig3} shows the calculated $k_T$ dependence of the HBT radii for central (0-10\%) Au+Au collisions at $\sqrt{s_\mathrm{NN}}=2.4 - 7.7$ GeV.
To clarify the influence of nuclear potentials on the pion interferometry, the radii are calculated with and without hadronic potentials (cascade: solid black lines with full squares, hard EoS: solid blue lines with full circles, soft EoS: solid pink lines) and compared with the experimental data \cite{Lisa:2005dd,HADES:2018gop,HADES:2019lek,E895:2000bxp,Okorokov:2013fya,STAR:2014shf,STAR:2020dav}.

One can clearly see that with increasing stiffness of the potential (i.e. stronger repulsion as function of density), the $R_{O}$ at large $k_T$ is driven down while the $R_{S}$ at small $k_T$ is pulled up at low beam energies.
This results in a decrease of the $R_{O}/R_{S}$ ratio as a function of $k_T$ and allows for a better description of the experimental data (can be seen in Fig.\ref{fig4} (a)).
This is due to the repulsive nature of the interactions which reflects the positive potential $V$ at large densities.
In addition, it is seen that both the values of the HBT-radii and their decrease with $k_T$ can be well reproduced by the calculations with a  hard Skyrme potential EoS at low beam energies.
The origin of the HBT-radii decreases with increasing transverse momentum has been discussed in many works, we refer the interested reader to related works for the details \cite{Li:2006gp,STAR:2004qya,STAR:2020dav,Fang:2022kru,Li:2006gb}.
At $\sqrt{s_\mathrm{NN}}=2.4$ GeV, also results using the soft Skyrme EoS ($K_{0}=200$ MeV) are added for comparison.
Compared to the results of simulations with a hard EoS, it can be seen that in simulations with a soft EoS, the $R_{O}$ will be increased while $R_{S}$ is decreased, and the $R_{O}/R_{S}$ ratio is increased consequently showcasing the general EoS dependence.
Based on the above results, we  conclude that a repulsive density dependent EoS will lead to a stronger phase-space correlation explaining the HBT time-related tensions \cite{Pratt:2008qv} and leads to a larger emission source.

\subsection{Effects of the EoS with phase transitions}

\begin{figure}[ht]
\begin{centering}
\includegraphics[width=0.5\textwidth]{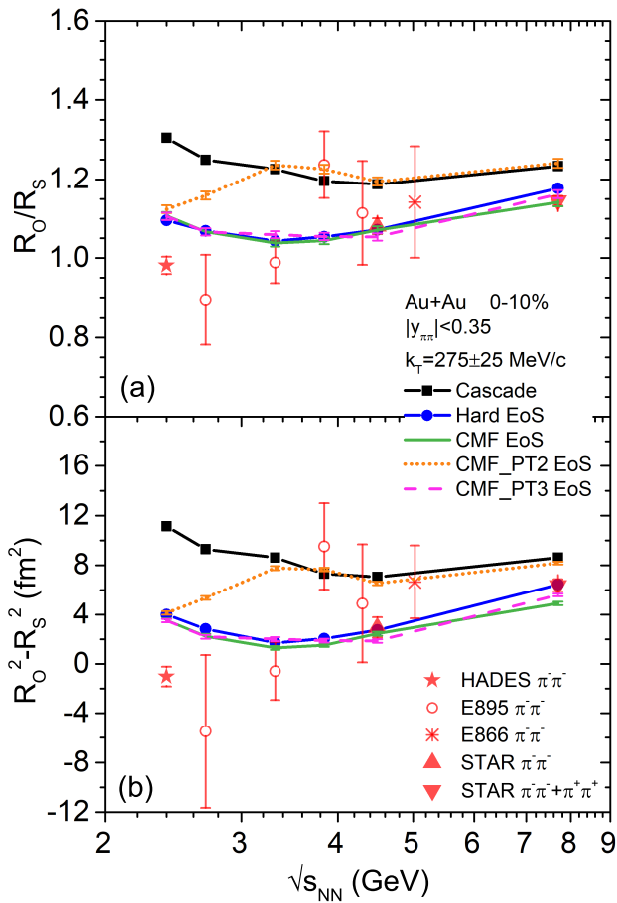}
\caption {\label{fig4}(Color online) The collision energy dependence of the $R_{O}/R_{S}$ (top panel) and the $R^{2}_{O}-R^{2}_{S}$ (bottom panel) extracted from freeze-out $\pi^{-}\pi^{-}$ without and with various EoS in central Au+Au collisions compared with the experimental data taken from Refs. \cite{Lisa:2005dd,HADES:2018gop,HADES:2019lek,E895:2000bxp,Okorokov:2013fya,STAR:2014shf,STAR:2020dav}.}
\end{centering}
\end{figure}

It has been argued that a long emission timescale $\Delta\tau$ may arise if the system evolves through a first-order phase transition, which stalls the expansion because the speed of sound vanishes. This should result in a strong increase of the $R_{O}$  compared with the $R_{S}$ \cite{Pratt:1986cc,Bertsch:1988db,Rischke:1996em}.
Thus, the difference $R^{2}_{O}-R^{2}_{S}$ and the ratio $R_{O}/R_{S}$ can provide information\footnote{It is noteworthy that only for a static (non-flowing) source, the emission time can be directly given by $\beta_{t}^{2}\Delta\tau^{2}=R_{O}^{2}-R_{S}^{2}$ \cite{Heinz:1996qu}, where $\beta_{t}=k_{T}/m_{T}$ is the transverse velocity of the emitted pions, for a flowing source, this relationship is unreliable and extracting timescales from $R_{O}^{2}-R_{S}^{2}$ becomes model dependent \cite{Lisa:2005dd,E895:2000bxp,Retiere:2003kf,STAR:2014shf}.}
on the emission duration, which might be extended if the system undergoes a phase transition.

Experimentally, the situation is unfortunately a bit unclear.
In the interesting energy region, the data has a rather large error bar and supports both interpretations \cite{Lacey:2014wqa,STAR:2014shf,HADES:2018gop}: a local maximum around $\sqrt{s_\mathrm{NN}}\approx 4$ GeV in the excitation functions of $R_{O}/R_{S}$ and $R^{2}_{O}-R^{2}_{S}$ or a smooth increase.

To obtain realistic quantitative predictions for the expected change of the emission time duration due to a phase transition, we employ the UrQMD model with a new equation of state including a phase transition.
This allows us to directly simulate the effect of a phase transition on  $R_{O}/R_{S}$ and $R^{2}_{O}-R^{2}_{S}$  in a consistent manner and pin down the previous qualitative predictions in a quantitatively realistic setup.
Fig. \ref{fig4} compares the collision energy dependence of $R_{O}/R_{S}$ (top panel) and $R^{2}_{O}-R^{2}_{S}$ (bottom panel) calculated with various EoS with a broad range of experimental data.
One observes that by considering the CMF EoS, the ratio $R_{O}/R_{S}$ and the square difference $R^{2}_{O}-R^{2}_{S}$ are pulled down in comparison to the cascade mode, and the present data can be qualitatively reproduced.
In this energy range the CMF EoS gives very similar results to the hard Skyrme EoS which also includes a strong repulsion leading to earlier pion emission. Generally the effects of the equation of state decrease with increasing collision energy.

Let us now turn to the EoS with a phase transition. Here we compare two CMF EoS, both including a phase transition are adopted.
The CMF$\_$PT2 EoS includes a phase transition at low baryon densities, while the CMF$\_$PT3 EoS includes a phase transition at higher baryon densities (cf. Fig. \ref{fig1}).
At the lowest energy ($\sqrt{s_\mathrm{NN}}=2.4$ GeV), the results calculated with all CMF EoS are similar as the EoS agree up to 2.5 saturation density.
As the collision energy increases, the calculated results of CMF$\_$PT2 EoS gradually increase compared to the standard CMF (or Hard/CMF$\_$PT3) EoS as expected for the appearance of a phase transition, interestingly and are similar to those with the cascade mode at $\sqrt{s_\mathrm{NN}}=3.3$ GeV ($E_{\rm{lab}}=4$ GeV/nucleon).
This is understood since the pure cascade mode can be considered a super soft EoS and therefore behaves similarly to a phase transition.
In addition, the results from simulations with the Hard EoS, CMF EoS and CMF$\_$PT3 EoS are close to each other in the whole energy region under investigation. In conjunction with Fig. \ref{fig1}, the above phenomenon can be well understood.
The baryon density of 0-10$\%$ central Au+Au collisions at $\sqrt{s_\mathrm{NN}}=2.4$ GeV is less than $3\rho_{0}$, and the density reaches about $5\rho_{0}$ for 0-10$\%$ central Au+Au collisions at $\sqrt{s_{\rm{NN}}}=7.7$ GeV ($E_{\rm{lab}}=29.7$ GeV/nucleon).
Thus, the HBT radii calculated with CMF$\_$PT2 are a result of the phase transition encountered for most collision energies while the transition in PT3 is never really reached, even for the highest collision energy.
Thus  using CMF$\_$PT3 shows no signal of the phase transition in the explored energy regime.
Our results indicate that the pion HBT radii parameters $R_{O}/R_{S}$ and $R^{2}_{O}-R^{2}_{S}$ are very sensitive to the EoS up to densities of 4-5 times saturation density only and are consistent with the absence of any strong softening due to a phase transition up to that point.

\subsection{Discussions}

To better understand the results with the different EoS let us discuss the freeze-out times and coordinates in the following section.

\begin{figure}[ht]
\begin{centering}
\includegraphics[width=0.5\textwidth]{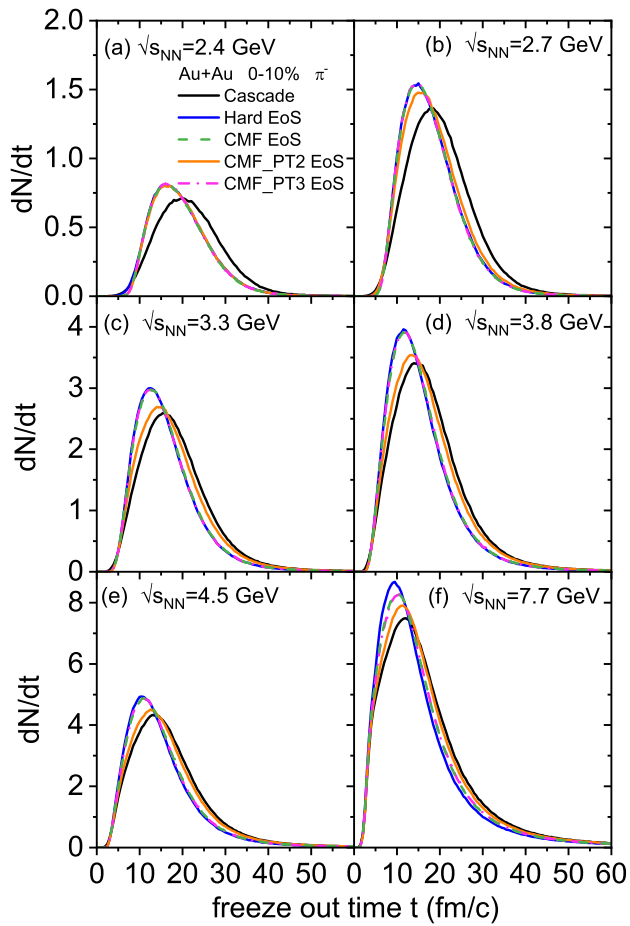}
\caption {\label{fig5}(Color online) Particle production yield as a function of freeze-out time for $\pi^{-}$ from 0–10\% Au+Au collisions. Calculations with the cascade mode are compared with the simulations with potentials.}
\end{centering}
\end{figure}

Fig. \ref{fig5} shows the freeze-out time distribution of the $\pi^{-}$ emission in central Au+Au collisions in the inspected energy region.
The results from the different equations of state are represented by various coloured lines, respectively.
It can be clearly seen that pions are mainly frozen-out in the time interval 5-25 fm/$c$, and that the pions are frozen-out earlier in case of a harder EoS.
In addition, at $\sqrt{s_\mathrm{NN}}=2.4$ GeV, the distributions of the results from all simulations with potentials are almost identical, and different from the distribution using cascade mode, simply because the EoS for such low densities is very similar for all density dependent potentials used.
As the energy increases, the distributions from simulations with hard, CMF and CMF$\_$PT3 EoS remain the same, while the distribution from simulations calculated with CMF$\_$PT2 EoS gradually approach that of the soft cascade calculations. At $\sqrt{s_\mathrm{NN}}=7.7$ GeV, owing to the CMF and CMF$\_$PT3 EoS being softer than the hard Skyrme potential and stiffer than the CMF$\_$PT2 EoS, the distributions from simulations with CMF and  CMF$\_$PT3 EoS lie between the distributions of simulations with hard EoS and CMF$\_$PT2 EoS.

\begin{figure}[ht]
\begin{centering}
\includegraphics[width=0.5\textwidth]{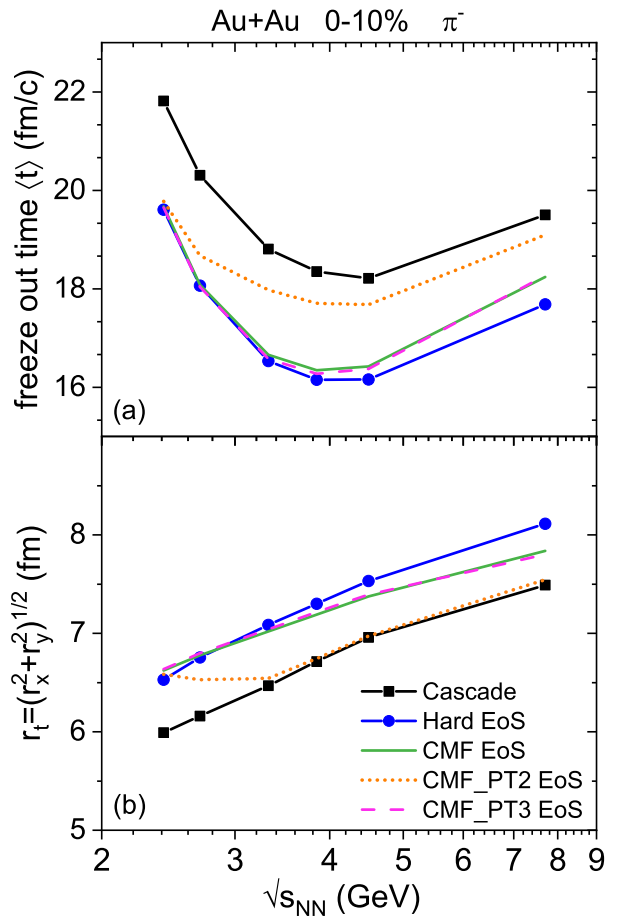}
\caption {\label{fig6}(Color online) Extracted average $\pi^{-}$ emission times $\langle t\rangle$ and transverse radii $r_{t}$ at freeze-out time as a function of collision energies depending on the EoS used.}
\end{centering}
\end{figure}

The mean values of the $\pi^{-}$ freeze-out time [panel (a)] and the transverse radii [panel (b)] are plotted in Fig. \ref{fig6} shown as different coloured lines with symbols.
It can be seen that the excitation function of the mean values of the freeze-out time shows a minimum at about $\sqrt{s_\mathrm{NN}}=4$ GeV for all calculations with and without phase transition.
This is likely due to the combination and competition of the resonance decay with the string excitation and fragmentation which leads to a change in the collision geometry towards higher collision energies.
In addition, one finds that the mean values of the freeze-out time (transverse radii) from hard equations of state are smaller (larger) than that of the softer ones.
This general behaviour is understood as a result of the larger pressure generated by the potentials, leading to a stronger expansion, consequently larger transverse radii and an earlier freeze-out time.

\section{Conclusions}\label{sec:4}
The UrQMD transport model was used, to systematically study the equation of state effects on the pion interferometry at  collision energies from $\sqrt{s_\mathrm{NN}}=2.4 - 7.7$ GeV.
To this aim, UrQMD was supplemented with novel equations of state based on the CMF model, including a phase transition at high baryon densities.

It was found that the HBT radii and the time-related ratio $R_{O}/R_{S}$ are weakly affected by the Coulomb potentials during the evolution of the system.
However, the source radii parameters ($R_{O}/R_{S}$ and $R^{2}_{O}-R^{2}_{S}$) where shown to be sensitive to the equation of state at densities up to 4-5 times nuclear saturation density.
The present experimental data, in the investigated energy region, can be qualitatively and quantitatively reproduced by simulations with an equation of state that shows stiff behaviour up to 4 times saturation density and a consecutive softening.

By comparing to the available HBT data we can exclude the existence of a strong phase transition for densities up to 4-5 times saturation density. The present study enables us to directly relate phase transition effects from pion HBT to e.g. flow observables predicted within the same approach \cite{Steinheimer:2022gqb}. Only this way, one can obtain a consistent picture of the high density equation of state of QCD from comparisons with experimental data.

Generally, the effects of the density dependent equation of state on the HBT radii are shown to decrease with increasing collision energy so that statements for higher densities are yet unreliable.

Unfortunately, in the most interesting energy region, the experimental data still shows substantial errors. To elucidate the details of the equation of state with HBT data, more theoretical works on understanding the uncertainty from the model are needed on the one hand, and highly accurate experimental data is desired on the other hand.

\begin{acknowledgements}
The work is supported in part by the National Natural Science Foundation of China (Nos. 11875125 and 12075085).
P. Li gratefully acknowledges the financial support from China Scholarship Council (No. 202106180053).
J. S. thanks the Samson AG for funding.
The authors are grateful to the C3S2 computing center in Huzhou University for calculation support.
\end{acknowledgements}




\begin{thebibliography}{100}

\bibitem{Pandav:2022xxx}
A.~Pandav, D.~Mallick and B.~Mohanty,
Prog. Part. Nucl. Phys. \textbf{125} (2022), 103960.

\bibitem{Aoki:2006we}
Y.~Aoki, G.~Endrodi, Z.~Fodor, S.~D.~Katz and K.~K.~Szabo,
Nature \textbf{443}, 675-678 (2006).

\bibitem{Bazavov:2011nk}
A.~Bazavov, T.~Bhattacharya, M.~Cheng, C.~DeTar, H.~T.~Ding, S.~Gottlieb, R.~Gupta, P.~Hegde, U.~M.~Heller and F.~Karsch, \textit{et al.}
Phys. Rev. D \textbf{85} (2012), 054503.


\bibitem{Braun-Munzinger:2015hba}
P.~Braun-Munzinger, V.~Koch, T.~Sch\"afer and J.~Stachel,
Phys. Rept. \textbf{621} (2016), 76-126.


\bibitem{Bzdak:2019pkr}
A.~Bzdak, S.~Esumi, V.~Koch, J.~Liao, M.~Stephanov and N.~Xu,
Phys. Rept. \textbf{853} (2020), 1-87.

\bibitem{Gunkel:2021oya}
P.~J.~Gunkel and C.~S.~Fischer,
Phys. Rev. D \textbf{104} (2021) 054022.

\bibitem{starnote0598}
Studying the Phase Diagram of QCD Matter at RHIC, (STAR Notes) SN0598,
02 June, 2014, https://drupal.star.bnl.gov/STAR/starnotes/public/sn0598.

\bibitem{STAR:2002eio}
K.~H.~Ackermann \textit{et al.} [STAR Collaboration],
Nucl. Instrum. Meth. A \textbf{499} (2003), 624-632.


\bibitem{Bratkovskaya:2004kv}
E.~L.~Bratkovskaya, M.~Bleicher, M.~Reiter, S.~Soff, H.~Stoecker, M.~van Leeuwen, S.~A.~Bass and W.~Cassing,
Phys. Rev. C \textbf{69} (2004), 054907.

\bibitem{Bratkovskaya:2013vx}
E.~L.~Bratkovskaya, J.~Aichelin, M.~Thomere, S.~Vogel and M.~Bleicher,
Phys. Rev. C \textbf{87} (2013), 064907.

\bibitem{Reichert:2021ljd}
T.~Reichert, A.~Elz, T.~Song, G.~Coci, M.~Winn, E.~Bratkovskaya, J.~Aichelin, J.~Steinheimer and M.~Bleicher,
J. Phys. G \textbf{49} (2022), 055108.

\bibitem{Bleicher:2022kcu}
M.~Bleicher and E.~Bratkovskaya,
Prog. Part. Nucl. Phys. \textbf{122} (2022), 103920.

\bibitem{TMEP:2022xjg}
H.~Wolter \textit{et al.} [TMEP Collaboration],
Prog. Part. Nucl. Phys. \textbf{125} (2022), 103962.

\bibitem{Bass:1999tu}
S.~A.~Bass, A.~Dumitru, M.~Bleicher, L.~Bravina, E.~Zabrodin, H.~Stoecker and W.~Greiner,
Phys. Rev. C \textbf{60} (1999), 021902.

\bibitem{Dumitru:1999sf}
A.~Dumitru, S.~A.~Bass, M.~Bleicher, H.~Stoecker and W.~Greiner,
Phys. Lett. B \textbf{460} (1999), 411-416.

\bibitem{Steinheimer:2007iy}
J.~Steinheimer, M.~Bleicher, H.~Petersen, S.~Schramm, H.~Stocker and D.~Zschiesche,
Phys. Rev. C \textbf{77} (2008), 034901.

\bibitem{Huovinen:2006jp}
P.~Huovinen and P.~V.~Ruuskanen,
Ann. Rev. Nucl. Part. Sci. \textbf{56} (2006), 163-206.

\bibitem{Petersen:2008dd}
H.~Petersen, J.~Steinheimer, G.~Burau, M.~Bleicher and H.~St\"ocker,
Phys. Rev. C \textbf{78} (2008), 044901.

\bibitem{Steinheimer:2011ea}
J.~Steinheimer, S.~Schramm and H.~Stocker,
Phys. Rev. C \textbf{84} (2011), 045208.

\bibitem{Shen:2020mgh}
C.~Shen and L.~Yan,
Nucl. Sci. Tech. \textbf{31} (2020), 122.

\bibitem{STAR:2020tga}
J.~Adam \textit{et al.} [STAR Collaboration],
Phys. Rev. Lett. \textbf{126} (2021), 092301.

\bibitem{STAR:2021fge}
M.~S.~Abdallah \textit{et al.} [STAR Collaboration],
Phys. Rev. Lett. \textbf{128} (2022), 202303.

\bibitem{NA49:2012ebu}
T.~Anticic \textit{et al.} [NA49 Collaboration],
Eur. Phys. J. C \textbf{75} (2015), 587.

\bibitem{Sun:2018jhg}
K.~J.~Sun, L.~W.~Chen, C.~M.~Ko, J.~Pu and Z.~Xu,
Phys. Lett. B \textbf{781} (2018), 499-504.

\bibitem{Lacey:2014wqa}
R.~A.~Lacey,
Phys. Rev. Lett. \textbf{114} (2015), 142301.

\bibitem{Bluhm:2020mpc}
M.~Bluhm, A.~Kalweit, M.~Nahrgang, M.~Arslandok, P.~Braun-Munzinger, S.~Floerchinger, E.~S.~Fraga, M.~Gazdzicki, C.~Hartnack and C.~Herold, \textit{et al.}
Nucl. Phys. A \textbf{1003} (2020), 122016.

\bibitem{HanburyBrown:1954amm}
R.~Hanbury Brown and R.~Q.~Twiss,
Phil. Mag. Ser. 7 \textbf{45} (1954), 663-682.

\bibitem{HanburyBrown:1956bqd}
R.~Hanbury Brown and R.~Q.~Twiss,
Nature \textbf{178} (1956), 1046-1048.

\bibitem{gglp} G. Goldhaber, S. Goldhaber,Won-Yong Lee, and A. Pais,
Phys. Rev. \textbf{120} 300, 1960.


\bibitem{Zajc:1984vb}
W.~A.~Zajc, J.~A.~Bistirlich, R.~R.~Bossingham, H.~R.~Bowman, C.~W.~Clawson, K.~M.~Crowe, K.~A.~Frankel, J.~G.~Ingersoll, J.~M.~Kurck and C.~J.~Martoff, \textit{et al.}
Phys. Rev. C \textbf{29} (1984), 2173-2187.

\bibitem{Pratt:1984su}
S.~Pratt,
Phys. Rev. Lett. \textbf{53} (1984), 1219-1221.

\bibitem{E895:2000bxp}
M.~A.~Lisa \textit{et al.} [E895 Collaboration],
Phys. Rev. Lett. \textbf{84} (2000), 2798-2802.

\bibitem{Lisa:2005dd}
M.~A.~Lisa, S.~Pratt, R.~Soltz and U.~Wiedemann,
Ann. Rev. Nucl. Part. Sci. \textbf{55} (2005), 357-402.

\bibitem{STAR:2014shf}
L.~Adamczyk \textit{et al.} [STAR Collaboration],
Phys. Rev. C \textbf{92} (2015), 014904.

\bibitem{Retiere:2003kf}
F.~Retiere and M.~A.~Lisa,
Phys. Rev. C \textbf{70} (2004), 044907.

\bibitem{Pratt:1986cc}
S.~Pratt,
Phys. Rev. D \textbf{33} (1986), 1314-1327.

\bibitem{Bertsch:1988db}
G.~Bertsch, M.~Gong and M.~Tohyama,
Phys. Rev. C \textbf{37} (1988), 1896-1900.

\bibitem{Rischke:1996em}
D.~H.~Rischke and M.~Gyulassy,
Nucl. Phys. A \textbf{608} (1996), 479-512.

\bibitem{Li:2008qm}
Q.~Li, J.~Steinheimer, H.~Petersen, M.~Bleicher and H.~Stocker,
Phys. Lett. B \textbf{674} (2009), 111-116.

\bibitem{Zhang:2017axr}
C.~J.~Zhang and J.~Xu,
Phys. Rev. C \textbf{96} (2017), 044907.

\bibitem{Batyuk:2017smw}
P.~Batyuk, I.~Karpenko, R.~Lednicky, L.~Malinina, K.~Mikhaylov, O.~Rogachevsky and D.~Wielanek,
Phys. Rev. C \textbf{96} (2017), 024911.

\bibitem{Alqahtani:2020daq}
M.~Alqahtani and M.~Strickland,
Phys. Rev. C \textbf{102} (2020), 064902.

\bibitem{OmanaKuttan:2022the}
M.~Omana Kuttan, A.~Motornenko, J.~Steinheimer, H.~Stoecker, Y.~Nara and M.~Bleicher,
Eur. Phys. J. C \textbf{82} (2022), 427

\bibitem{Steinheimer:2022gqb}
J.~Steinheimer, A.~Motornenko, A.~Sorensen, Y.~Nara, V.~Koch and M.~Bleicher,
[arXiv:2208.12091 [nucl-th]].

\bibitem{Steinheimer:2018rnd}
J.~Steinheimer, Y.~Wang, A.~Mukherjee, Y.~Ye, C.~Guo, Q.~Li and H.~Stoecker,
Phys. Lett. B \textbf{785} (2018), 40-45.

\bibitem{Ye:2020lrc}
Y.~Ye, Y.~Wang, Q.~Li, D.~Lu and F.~Wang,
Phys. Rev. C \textbf{101} (2020), 034915.

\bibitem{Bass:1998ca}
  S.~A.~Bass, M.~Belkacem, M.~Bleicher, M.~Brandstetter, L.~Bravina, C.~Ernst, L.~Gerland and M.~Hofmann {\it et al.},
  Prog.\ Part.\ Nucl.\ Phys.\  {\bf 41}, (1998), 255.

\bibitem{Bleicher:1999xi}
  M.~Bleicher, E.~Zabrodin, C.~Spieles, S.~A.~Bass, C.~Ernst, S.~Soff, L.~Bravina, M.~Belkacem {\it et al.},
  J.\ Phys.\ G {\bf 25}, (1999), 1859.


\bibitem{Li:2010ie}
Q.~Li and Z.~Li,
Mod. Phys. Lett. A \textbf{27} (2012), 1250004.

\bibitem{Hillmann:2018nmd}
P.~Hillmann, J.~Steinheimer and M.~Bleicher,
J. Phys. G \textbf{45} (2018), 085101.

\bibitem{Hillmann:2019wlt}
P.~Hillmann, J.~Steinheimer, T.~Reichert, V.~Gaebel, M.~Bleicher, S.~Sombun, C.~Herold and A.~Limphirat,
J. Phys. G \textbf{47} (2020), 055101.

\bibitem{Li:2005gfa}
Q.~F.~Li, Z.~X.~Li, S.~Soff, M.~Bleicher and H.~Stoecker,
J. Phys. G \textbf{32} (2006), 151-164.


\bibitem{Steinheimer:2009nn}
J.~Steinheimer, V.~Dexheimer, H.~Petersen, M.~Bleicher, S.~Schramm and H.~Stoecker,
Phys. Rev. C \textbf{81} (2010), 044913,

\bibitem{Motornenko:2019arp}
A.~Motornenko, J.~Steinheimer, V.~Vovchenko, S.~Schramm and H.~Stoecker,
Phys. Rev. C \textbf{101} (2020), 034904.

\bibitem{Steinheimer:2010ib}
J.~Steinheimer, S.~Schramm and H.~Stocker,
J. Phys. G \textbf{38} (2011), 035001.

\bibitem{Mukherjee:2016nhb}
A.~Mukherjee, J.~Steinheimer and S.~Schramm,
Phys. Rev. C \textbf{96} (2017), 025205.

\bibitem{Motornenko:2020yme}
A.~Motornenko, S.~Pal, A.~Bhattacharyya, J.~Steinheimer and H.~Stoecker,
Phys. Rev. C \textbf{103} (2021), 054908.

\bibitem{Jakobus:2022ucs}
P.~Jakobus, B.~Mueller, A.~Heger, A.~Motornenko, J.~Steinheimer and H.~Stoecker,
[arXiv:2204.10397 [astro-ph.HE]].

\bibitem{Most:2022wgo}
E.~R.~Most, A.~Motornenko, J.~Steinheimer, V.~Dexheimer, M.~Hanauske, L.~Rezzolla and H.~Stoecker,
[arXiv:2201.13150 [nucl-th]].

\bibitem{Seck:2020qbx}
F.~Seck, T.~Galatyuk, A.~Mukherjee, R.~Rapp, J.~Steinheimer and J.~Stroth,
Phys. Rev. C \textbf{106} (2022), 014904.

\bibitem{CRAB}
https://web.pa.msu.edu/people/pratts/freecodes/crab/home.html

\bibitem{Wong:2007hx}
C.~Y.~Wong and W.~N.~Zhang,
Phys. Rev. C \textbf{76} (2007), 034905.


\bibitem{Sinyukov:1998fc}
Y.~Sinyukov, R.~Lednicky, S.~V.~Akkelin, J.~Pluta and B.~Erazmus,
Phys. Lett. B \textbf{432} (1998), 248-257,

\bibitem{HADES:2018gop}
J.~Adamczewski-Musch \textit{et al.} [HADES Collaboration],
Phys. Lett. B \textbf{795} (2019), 446-451.

\bibitem{HADES:2019lek}
J.~Adamczewski-Musch \textit{et al.} [HADES Collaboration],
Eur. Phys. J. A \textbf{56} (2020), 140.


\bibitem{Okorokov:2013fya}
V.~A.~Okorokov,
[arXiv:1312.4269 [nucl-ex]].

\bibitem{STAR:2020dav}
J.~Adam \textit{et al.} [STAR Collaboration],
Phys. Rev. C \textbf{103} (2021), 034908.

\bibitem{STAR:2004qya}
J.~Adams \textit{et al.} [STAR Collaboration],
Phys. Rev. C \textbf{71} (2005), 044906.

\bibitem{Li:2006gb}
Q.~Li, M.~Bleicher, X.~Zhu and H.~Stoecker,
J. Phys. G \textbf{33} (2007), 537-548.

\bibitem{Li:2006gp}
Q.~Li, M.~Bleicher and H.~Stoecker,
Phys. Rev. C \textbf{73} (2006), 064908.

\bibitem{Fang:2022kru}
L.~M.~Fang, Y.~G.~Ma and S.~Zhang,
Eur. Phys. J. A \textbf{58} (2022), 81.

\bibitem{Pratt:2008qv}
S.~Pratt,
Phys. Rev. Lett. \textbf{102} (2009), 232301.


\bibitem{Heinz:1996qu}
U.~W.~Heinz, B.~Tomasik, U.~A.~Wiedemann and Y.~F.~Wu,
Phys. Lett. B \textbf{382} (1996), 181-188.
\end{thebibliography}
\end{document}